# Physical electrostatics of small field emitter arrays/clusters

Richard G. Forbes[a]

*Advanced Technology Institute & Department of Electrical and Electronic Engineering, University of Surrey, Guildford, Surrey GU2 7XH, UK*

This paper aims to improve qualitative understanding of electrostatic influences on apex field enhancement factors (AFEFs) for small field emitter arrays/clusters. Using the "floating sphere at emitter-plate potential" (FSEPP) model, it re-examines the electrostatics and mathematics of three simple systems of identical post-like emitters. For the isolated emitter, various approaches are noted. An adequate approximation is to consider only the effects of sphere charges and (for significantly separated emitters) image charges. For the 2-emitter system, formulas are found for charge-transfer ("charge-blunting") effects and neighbor-field effects, for widely spaced and for "sufficiently closely spaced" emitters. Mutual charge-blunting is always the dominant effect, with a related (negative) fractional AFEF-change $\delta_{\mathrm{two}}$. For sufficiently small emitter spacing $c$, $|\delta_{\mathrm{two}}|$ varies approximately as $1/c$; for large spacing, $|\delta_{\mathrm{two}}|$ decreases as $1/c^3$. In a 3-emitter equispaced linear array, differential charge-blunting and differential neighbor-field effects occur, but differential charge-blunting effects are dominant, and cause the "exposed" outer emitters to have higher AFEF ($\gamma_0$) than the central emitter ($\gamma_1$). Formulas are found for the exposure ratio $\varXi=\gamma_0/\gamma_1$, for large and for sufficiently small separations. The FSEPP model for an isolated emitter has accuracy around 30%. Line-charge models (LCMs) are an alternative, but an apparent difficulty with recent LCM implementations is identified. Better descriptions of array electrostatics may involve developing good fitting equations for AFEFs derived from accurate numerical solution of Laplace's equation, perhaps with equation form(s) guided qualitatively by FSEPP-model results. In existing fitting formulas, the AFEF-reduction decreases exponentially as $c$ increases, which is different from the FSEPP-model formulas. This discrepancy needs to be investigated, using systematic Laplace-based simulations and appropriate results analysis. FSEPP models might provide a useful provisional guide to the qualitative behaviour of small field emitter clusters larger than those investigated here.

## I. INTRODUCTION

Currently, there is significant technological interest in large-area field electron emitters (LAFEs), especially in LAFEs based on arrays of post-like carbon nanotubes or nanofibres. Recent reviews discuss potential applications.[1-3]

---

[a] Electronic mail: r.forbes@trinity.cantab.net

An important LAFE characterization parameter is the true (electrostatic) macroscopic field enhancement factor (FEF) defined below. Provided the emission situation is orthodox[4], a FEF-value can be validly extracted[5] from a Fowler-Nordheim plot of LAFE emission current or average current density—although interpretation may be problematic if individual emitters are not all similar or have irregular apexes. If emission is not orthodox, then the FEF-value found by orthodox or elementary data-analysis may be spuriously high[4], but a rough estimate of the true FEF can be found by phenomenological adjustment[5].

Because the widely-used symbol $\beta$ has multiple meanings in field electron emission (FE) theory, and confusion has sometimes occurred, the author prefers $\gamma$ as the basic symbol for true (electrostatic) FEFs, and prefers the short name "FEF".

There has been much interest in predicting FEFs, especially apex values for individual emitters. Simplifying assumptions often made, and made here, are: (a) LAFEs can be modelled as a set of post-like emitters; (b) each emitter stands upright on an "emitter plate" that is one of a pair of parallel, planar conducting plates separated by a distance $d_{sep}$ that is very large in comparison with all emitter heights; (c) the detailed structure of the emitter apex, and related effects, can initially be disregarded; (d) an emitter can be treated as a cylindrically symmetrical conducting post with a smooth classical surface (which means the local surface electrostatic field will have greatest absolute magnitude at the post apex); and (e) work-function variations can be disregarded, by assuming that all emitter and plate surfaces have the same uniform work function $\phi$ (which means the local emission current density has highest magnitude at the post apex).

There was early interest in the apex FEF $\gamma_{one}$ for an isolated conducting post, often treated as a cylinder with a hemispherical cap. No exact analytical solution is known. Existing treatments are either approximate analytical solutions, or are numerical—based on assumed or optimised charge distributions or numerical solution of Laplace's equation. References 6 and 7 discuss or note work prior to 2004; a more detailed 2011 treatment[8] ("ZPCL") gives more recent references; Refs. 9 and 10 are other papers of particular interest. The simplest useful approximations use the "floating sphere at emitter-plate potential (FSEPP)" model introduced into FE by Gomer[11] and then used by Vibrans[12,13]

and Beckey et al.[14]

There is also interest in predicting FEF-values for emitters in arrays/clusters. The general problem, where emitters have different shapes, is difficult. Thus, much work has looked at arrays/clusters of geometrically identical emitters.

Infinite regular arrays have been treated by ZPCL, using the FSEPP model, and very large regular arrays by several groups, either numerically or by using the FSEPP model (see ZPCL for references to both these), or by using a line-charge model (e.g., Ref. 15). The apex FEF at each emitter is changed from the value $\gamma_{one}$, due to the electrostatic influence of the other emitters. This effect is usually called "screening" or "mutual screening" or "shielding", but these names are not informative. As shown below, several distinguishable effects are involved. The overall set of effects is perhaps best described by the term *electrostatic influence*.

With infinite regular arrays, or a pair of identical emitters, two effects operate. The first is a *charge-distribution effect*: as the emitters are brought closer together, charge is forced back from the emitters onto the emitter plate, due to the laws of electron thermodynamics, and this leads to a reduction in apex FEF. The second is a *neighbor-field effect*: the total field at the apex of a given emitter contains contributions due to the charges representing the other emitter(s): this leads to apex-FEF increase if two emitters are sufficiently close, but to apex-FEF decrease for larger separations.

Finite arrays/clusters are of particular interest, because FEF-modification effects operate differently on different emitters (because their geometrical environments are different). Laplace-based numerical treatments (e.g., Refs 16-18) show that "exposed" emitters near array edges and corners have higher apex FEFs. This seems equivalent to the well-known effect, for solid bodies, that the magnitudes of field and surface charge density are highest where the body is geometrically sharpest.

In the arrays, emitters in exposed positions have higher tip currents. As Harris et al.[19] point out, these tip-current variations can have unwanted technological consequences. To deal with these, fuller understanding of why exposed emitters have high apex FEFs may be helpful. The author's perception is that there are two possible causes, namely (i) *differential charge distribution effects* (i.e., exposed emitters carry higher-than-average charge-magnitude), and/or (ii) *differential neighbor-field effects* (fewer neighbors are there to contribute to the apex field of an exposed emitter).

The possibility of neighbor-field effects is recognised in line-charge models (e.g., Ref. 15), but the author is not aware of any clear analytical treatment of differential charge-distribution effects in small emitter arrays. FSEPP model treatments do not normally take neighbour-field effects into account. The aims of this paper are, first, a "demonstration of method" of one way to analyse differential charge-distribution effects, and, second, what seems to be a first investigation of whether charge-distribution effects or neighbour-field effects have more influence on apex field modification. For transparency, a simple model (the FSEPP model) is applied to the simplest array in which differential effects occurs, namely a linear equispaced array of three geometrically identical emitters. A further aim is to discuss the background physics more completely then previously.

It needs emphasising that this paper does not aim to find accurate FEF values (this is best done via numerical solution of Laplace's equation), or to find complete answers. Rather, the aim is to gain additional physical understanding of issues and trends in the electrostatics of finite emitter arrays/ clusters, in so far as FSEPP models allow this, and indicate a route to future progress.

The paper's structure is as follows. Section II sets out the underlying physics. Section III looks again at an isolated emitter, in order to justify approximations used later. Section IV looks at the interaction of two identical emitters, to assess the physical electrostatic influences that operate in this case and for infinite regular arrays. Section V looks at the case of three equispaced identical emitters, which in addition exhibits differential effects. Section VI provides discussion. The paper uses what is now called[20,21] the "International System of Quantities (ISQ)" (i.e., the electric constant $\varepsilon_0$ appears in Coulomb's law).

## II. PRINCIPLES BEHIND THE ELECTROSTATIC MODELLING OF FIELD EMITTERS

### A. Conventions relating to "electric field"

The simplest method (used here) of discussing electrostatic problems in FE contexts is to use conventional electrostatics, in which: the symbol $\Phi$ denotes conventional electrostatic potential; the symbol $\boldsymbol{E}$ denotes conventional electrostatic field [$\boldsymbol{E} = -\mathbf{grad}\,\Phi$]; and the symbol $E$ denotes the *signed* magnitude of a conventional electrostatic field or field component (e.g., a component normal to the emitter surface). Distance $z$ is measured from the emitter plate, towards the opposing plate.

The local surface electrostatic field is denoted by $E_L$, and its apex value by $E_a$. Such fields are negative for field electron emitters. Thus, this convention requires that the absolute magnitude ($|E_L|$ or $|E_a|$) be used in Fowler-Nordheim-type equations (or that a different symbol, normally $F$, be used to denote $|E|$), and requires use of terms such as "higher-magnitude field".

This convention is different from the "electron emission convention" implicitly used in much FE recent literature, where the symbol $E$ is a positive quantity that denotes the *absolute* magnitude of a negative electrostatic field or field component, i.e. the quantity denoted here by $|E|$.

For a field electron emitter, the charge $q$ placed at the centre of a floating sphere is negative in value, but the theory is algebraically valid irrespective of whether $q$ (and hence $E_a$) are positive or negative. Derived FEF-values are, of course, positive in both cases. The author's view is that arguments about field enhancement are often easier to follow if emitters are thought of as positively charged. Care has been taken to make the text polarity independent.

Note that the convention used here, of denoting the charge at the centre of a sphere by $q$, is different from that used in some published papers, which denote this charge by $-q$.

**B. Definitions of field enhancement factor**

In parallel-plane-plate geometry, the *macroscopic* electrostatic field $E_M$ is given by

$$E_M = -\Delta\Phi / d_{sep} \approx -V_p / d_{sep}, \qquad (1)$$

where $\Delta\Phi$ [$=\Phi_c-\Phi_e$] is the difference in electrostatic potential between (points just outside the surfaces of) the counter-electrode ("c") and emitter ("e") plates, and $V_p$ is the corresponding voltage between these plates. When all surfaces are allocated the same local work function, $\Delta\Phi$ is numerically equal to $V_p$. In reality, local work-functions are not all equal, but errors are small in nearly all practical cases. Near-universal practice is to take $E_M = -V_p/d_{sep}$.

For a post-shaped emitter, a true (electrostatic) *macroscopic FEF* $\gamma_{ML}$ [$=E_L/E_M$] can be defined at any point on the emitter surface, but usually interest is in the post apex FEF $\gamma_{Ma}$ defined by

$$\gamma_{\text{Ma}} \equiv E_{\text{a}} / E_{\text{M}} \approx -E_{\text{a}} d_{\text{sep}} / V_{\text{p}} . \tag{2}$$

Provided $d_{\text{sep}}$ is very much greater than the emitter height (preferably at least five times the height), $\gamma_{\text{Ma}}$ is not a significant function of $d_{\text{sep}}$, but depends only on how the emitter shape affects the electrostatics. In this case, $\gamma_{\text{Ma}}$ is a parameter that characterizes the "sharpness" of the emitter alone (rather than the combined electrostatic behaviour of the emitter and counter-electrode).

With LAFEs, experimental interest is usually in the apex FEF for the most strongly emitting individual emitter, as this is how the theory for LAFE emission current and for LAFE FN plots is conventionally set up (e.g., see Ref. 5).

An alternative (but less useful) FEF-like parameter is the "true (electrostatic) *gap FEF*". Its value $\gamma_{\text{Ga}}$ for the apex of an individual emitter is defined by

$$\gamma_{\text{Ga}} \equiv E_{\text{a}} / E_{\text{G}} \approx -E_{\text{a}} d_{\text{gap}} / V_{\text{gap}} , \tag{3}$$

where $V_{\text{gap}}$ is the voltage between the counter-electrode (which may be an "anode probe") and the emitter, $d_{\text{gap}}$ is the length of the gap between them, and the *gap field* $E_{\text{G}}$ is given via eq. (3). In general, the parameter $\gamma_{\text{Ga}}$ depends on the whole system geometry, and specifically on the gap length $d_{\text{gap}}$. Current-voltage plots and FN plots may exhibit "shift" effects that depend on $d_{\text{gap}}$, and the extracted value of $\gamma_{\text{Ga}}$ may also depend on $d_{\text{gap}}$. Thus, although $\gamma_{\text{Ga}}$ may be useful in comparative studies of different emitters in arrays (e.g., Refs 22,23), and in other contexts where an anode-probe is used (e.g., Ref. 24), it is less useful than $\gamma_{\text{Ma}}$ as a characterization parameter for the sharpness of an individual emitter or an array.

The FEFs discussed above are called "true (electrostatic) FEFs" because they are determined only by the *electrostatics* of the geometry concerned (i.e., the situation that exists experimentally in the absence of significant current flow). This is to distinguish them from the slope characterization parameters ("apparent FEFs") derived from FN-plot analysis of measured current and voltage, which may be partly determined by the electrical characteristics of the current path between the emitter tip and the high voltage supply[4,5,25], or by other complicating factors[4]. When the emission situation is not "orthodox", apparent FEFs may be much greater than the true FEFs, perhaps as much as 100 times greater in the worst cases.[4]

In what follows, this paper will be interested only in the apex values of true macroscopic FEFs, and the suffix "M" will now be dropped.

**C. The condition for electrical equilibrium**

A classical electron conductor is in *internal electrical equilibrium* if the current density is zero at all internal points. The necessary thermodynamic condition is that the appropriately defined chemical potential $\mu$ for electrons be the same everywhere in the conductor. Fowler and Guggenheim[26] showed long ago that, for a free-electron conductor, the local $\mu$-value can be identified with the local electron Fermi level. Thus, the condition for internal electrical equilibrium is that the Fermi level be the same at all internal points. It is widely assumed that this condition applies (or applies adequately) to all conductors that have electrons as the predominant charge-carriers.

Electrostatic analyses of electron conductors usually disregard across-surface variations in work-function, and this is done here. This is not physically realistic in the absolute scale of things, but avoids significant complications. It allows the electrostatic potential $\Phi$ to be taken the same at all points "immediately outside" the surface of a classical conductor in internal electrical equilibrium. Without further loss of generality, one can allocate the value "zero" to this common value of $\Phi$.

The FSEPP model of a field emitter, and similar models, involve two stages. The first stage finds a configuration of point charges and dipoles that (to an adequate degree of mathematical approximation) can represent the emitter electrostatics by satisfying certain conditions relating to electrostatic potential $\Phi$. The second stage uses these model charges and dipoles to find the values of relevant fields, in particular the fields and FEFs at the apex(es) of the floating sphere(s) of interest. The dominant contribution to the apex field and FEF for a given sphere is that due to the point charge placed at its centre.

In the first stage, the most important requirement is that the electrostatic potentials at the apex(es) of each floating sphere be equal to the electrostatic potential just outside the emitter plate. In modelling, this potential is normally set equal to zero. Thus, this *pre-eminent requirement* becomes that the apex electrostatic potentials, in vacuum immediately outside the apex(es) of each floating sphere, be zero.

In the case of an isolated emitter, an arrangement of charges and dipoles is found (in particular, a charge $q_{\text{one}}$ is placed at the sphere centre) that adequately satisfies the pre-eminent requirement and (usually) one other electrostatic condition. Various mathematical approximations are possible; these give rise to various approaches and mathematical formulae, as discussed below. The primary physical parameters of interest are the *sphere radius r* and the *sphere elevation ℓ* (i.e., the perpendicular distance of the sphere centre from the emitter plate—see Fig. 1). When $\ell/r \gg 1$ (which is always true for practical emitters), then the various approaches all generate formulae close to the simplified result $\gamma_{\text{one}} \sim \ell/r$.

For clarity, the symbol $h$ is avoided in this work, as some papers use it to mean $\ell$ and others to mean $(\ell+r)$.

### D. Overview of the physics of FSEPP models

Due to the inherent linearity in basic electrostatics, the values of all charges and dipoles used in FSEPP models scale with the value of the macroscopic field $E_{\text{M}}$. Consequently, when considering system-geometry effects, one needs to consider how these influence values of $q_k/E_{\text{M}}$, where $q_k$ is the *k*th model charge of interest.

For an isolated emitter as modelled by a sphere of given radius, the positive quantity $(q_{\text{one}}/E_{\text{M}})$, increases as the elevation $\ell$ increases, as discussed elsewhere[6-15] and also below. This effect (increase in $q_{\text{one}}/E_{\text{M}}$, or more generally in $q_n/E_{\text{M}}$ where $q_n$ is the charge at the centre of sphere "n"), is called here *charge-sharpening*. Reduction in $q_n/E_{\text{M}}$ is called here *charge-blunting*. Both effects involve charge transfer between the sphere and the emitter plate.

The contributions $E_{q,n}$ and $E_{p,n}$ to the field at the apex of sphere "n", due to a point charge $q_n$ and point dipole $p_n$ at the sphere centre, are related to its radius $r_n$ by the usual formulas

$$E_{q,n} = q_n / 4\pi\varepsilon_0 r_n^2, \qquad E_{p,n} = 2p_n / 4\pi\varepsilon_0 r_n^3. \tag{4}$$

Clearly: (i) changes in $q_n$ and $p_n$, however caused, lead to changes in the related fields, and hence change in the apex FEF $\gamma_n$ for sphere "n"; and (ii) changes in $E_{q,n}$, $E_{p,n}$ and $\gamma_n$ can also result from a change in the given value of sphere radius $r_n$.

With two identical emitters, for later convenience labelled "0" and "n", one needs to consider what happens when the emitters are brought closer together. In this case, the charge distribution (point charges and dipoles) being used to represent emitter "n" will influence the total electrostatic potential $\Phi_{t,0}$ at the apex of emitter "0", and tend to cause $\Phi_{t,0}$ to change away from zero. The pre-eminent requirement above means that $\Phi_{t,0}$ must be kept at zero, and this is achieved by charge-blunting. This can alternatively be described as the induction, by the charge at the centre of sphere "n", of image charge (of the opposite sign) in sphere "0".

Emitter "0" has an equivalent effect on emitter "n", and the complete problem has to be solved self-consistently. The outcome is apex-FEF reduction for both emitters, due to *mutual charge-blunting*, involving charge transfer from the spheres to the emitter plate. This is an *indirect effect* of the proximity between "0" and "n".

Detailed calculations of the apex FEF for emitter "0" also need to include contributions due to: (i) the applied macroscopic field; (ii) other components (point charges and dipoles) of the charge distribution used to represent emitter "0"; and (iii) components (point charges and dipoles) of the charge distribution used to represent emitter "n". I refer to (iii) as the *neighbor-field effect;* this is the *direct effect* of emitter "n" on the calculation of apex field and FEF for emitter "0". Issues relating to the sign of the neighbor-field (NF) effect, and to the relative sizes of the neighbor-field and charge-blunting (CB) effects are discussed below.

With *infinite* regular arrays, the physical effects that can occur are similar to those occurring with two emitters, but the detailed mathematics includes sums taken over all emitter pairs.

With *finite* regular arrays/clusters, extra effects come into play, because the geometrical environments are not equivalent for all emitters. These extra effects are differences in the degree of charge-blunting and differences in neighbor-field effects, as between different emitters (or as between different classes of geometrically equivalent emitter). The former can alternatively be seen as partially involving charge transfer between geometrically non-equivalent emitters.

Mathematically, the main difference between infinite and finite regular arrays is as follows. Depending on the approximation used, solution for the charge-blunting effect in an infinite array requires solution of a single equation or two simultaneous equations. Solution for the charge-blunting effect in a finite array requires solution of a set of simultaneous equations; in the simplest approaches the number of equations equals the number of geometrically non-equivalent classes of emitter. Section V illustrates this for the linear array of three equispaced identical emitters, where there are two emitter classes.

For large finite arrays, there is also a separate charge-redistribution effect, which makes the apex FEFs for emitters near the centre of the array depend on the array size. This effect is not discussed in the present paper.

### III. ANALYSIS OF A ISOLATED EMITTER

As background to later Sections, it is useful to revisit the mathematics of the FSEPP model for an isolated emitter, as illustrated in Fig. 1. The resulting charge strengths, dipole strengths, and apex potential contributions are shown in Table I, for each of five different mathematical approaches (I to V) defined below; the resulting apex field and FEF contributions are shown in Table II. To keep track of small terms, and have consistency with ZPCL's treatment, $\eta$ is used to denote the ratio $r/\ell$. For a practical emitter with high apex FEF $\gamma_{one}$, the parameter $\eta \sim 1/\gamma_{one}$; $\eta$ is very much less than unity (typically of order 0.01 or less).

#### A. Analysis based on placing charges and dipoles at sphere centres

Approaches I to V limit the possible positions for point charges and dipoles to the centre of the floating sphere and the related image position. These are not exact physical treatment of a FSEPP model, but this keeps the mathematics straightforward. As shown below, this procedure yields good approximations. Thus, construction of an "adequately self-consistent" charge distribution proceeds via the following sequence of steps (e.g., Ref. 6).

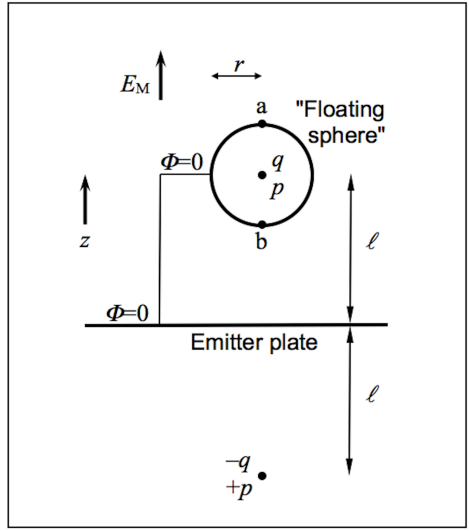

FIG. 1. To illustrate the "floating sphere at emitter plate potential" (FSEPP) model of a post-like field emitter, and related parameters. The counter-electrode (not shown) is at a distance $d_{\text{sep}}$ from the emitter plate that is very very much greater than $\ell$. The diagram is not to scale; for practical emitters the ratio ($\ell/r$) is typically around 100 or more.

(1) At the sphere apex, the macroscopic field (acting by itself) creates a potential contribution $\{-(\ell+r)E_M\}$ $[=-(1+\eta)\ell E_M]$.

(2) In addition, the field $E_M$ polarizes the sphere, inducing a surface charge distribution. Its effects, at and outside the sphere surface, are simulated by placing a point dipole (the *sphere dipole*) at the sphere centre, with its direction parallel to $E_M$. The dipole strength $p^0$ must be such that the potential difference $\Delta\Phi$ $[=\Phi_a-\Phi_b]$ between the sphere apex "a" and the diametrically opposed point "b", due to both field $E_M$ and the dipole, is zero. This requires that

$$\Delta\Phi = 2p^0/4\pi\varepsilon_0 r^2 - 2rE_M = 0, \tag{5}$$

i.e., $p^0/4\pi\varepsilon_0=r^3 E_M$. At the sphere apex, this dipole yields a potential contribution $p^0/4\pi\varepsilon_0 r^2$ $[=rE_M]$ that adds to that directly created by $E_M$, to give a total of $-\ell E_M$. More generally, the dipole ensures that the potential at all points on the sphere surface equals $-\ell E_M$.

(3) To reduce the sphere surface potential to the emitter plate potential (zero), a point charge (the *sphere charge*) is placed at the sphere centre. This needs to have strength $q^0$ given via

$$q^0/4\pi\varepsilon_0 r = \ell E_M, \tag{6}$$

i.e., $q^0/4\pi\varepsilon_0 = r\ell E_M$. Thus, the combined effects of the macroscopic field, the charge and the dipole create an apex potential $\Phi_a$ equal to 0, and an apex FEF $\gamma^{II} = (\ell/r)+3 = \eta^{-1}+3$, as shown in Tables I and II. This is the on-axis value of a well-known older result[27] and is equivalent to Gomer's "case 2" result[11] (though his discussion is formulated differently), and to eq. (7) in Ref. 12.

(4) A side-effect of introducing the sphere charge and (to a lesser extent) the sphere dipole is to create unwanted potential variation across the emitter plate. This is eliminated by introducing an image charge and image dipole located at distance $\ell$ behind the plate.

In principle, a similar procedure is needed for the counter-electrode plate, as (for example) in Refs 28 and 29; however, if $d_{sep} >>> \ell$, then the resulting corrections are very small and can be neglected.

To allow the option of different mathematical approximations, we now change to denoting the sphere charge by $q$ and the sphere dipole strength by $p$, and formulate "adequately self-consistent" equations for these.

(5) At the emitter apex, the image charge will make a potential contribution of $\{-q/4\pi\varepsilon_0(2\ell+r)\}$ and the image dipole a contribution of $\{-p/4\pi\varepsilon_0(2\ell+r)^2\}$. The image-charge will also alter the value of $\Delta\Phi$ in eq. (5), by an amount given by the first two terms in eq. (8). The change $\delta_{id}$ in $\Delta\Phi$ caused by the image dipole is very small, so an explicit expression is not given here. Thus, adjusted values for $q$ and $p$ are found via the simultaneous equations

$$\Phi_a = (q/4\pi\varepsilon_0)[r^{-1} - (2\ell+r)^{-1}] - (\ell+r)E_M + p/4\pi\varepsilon_0 r^2 + p/\{4\pi\varepsilon_0(2\ell+r)^2\} = 0, \tag{7}$$

$$\Delta\Phi = (q/4\pi\varepsilon_0)[(2\ell-r)^{-1} - (2\ell+r)^{-1}] - 2rE_M + 2p/4\pi\varepsilon_0 r^2 + \delta_{id} = 0 \tag{8}$$

The term $q/4\pi\varepsilon_0 r$ in eq. (7) is the sphere-charge term; the other terms involving $q$ in eqs (7) and (8) are image-charge terms. Eqs (7) and (8) can be regarded as a model analysis that is "adequately close to self-consistent" (if $d_{sep} \ggg \ell$).

**B. The basic mathematical approximations**

Equations (7) and (8) can be solved self-consistently as they stand, or in various approximations. In all cases, the macroscopic-field and sphere-charge terms are kept.

The simplest approach (I) uses a version of eq. (7) containing only these terms. This yields the basic result $\gamma^I = (\ell/r) + 2 = \eta^{-1} + 2$.

The next simplest approach (II), already described above, additionally keeps the sphere-dipole terms in both equations, but disregards all image terms.

Approach III, used in Section C of ZPCL, and also in Section 2.2 of Ref. 6 (but analysed in a different way there), adds to Approach II the image-charge term in eq. (7). Equation (8) still yields the sphere-dipole strength as $p = p^0$, and eq. (7) becomes

$$(q/4\pi\varepsilon_0)[r^{-1} - (2\ell + r)^{-1}] = \ell E_M. \tag{9}$$

In Approach III, the strengths $q$ and $p$, and all contributing potential terms in Table I, are mathematically *exact*. The final result for $\gamma^{III}$ is also given as eq. (11) in ZPCL. Formula (12) in Ref. 6 is an approximated version.

Approach IV (also used in Section D of ZPCL) adds to Approach III the image-charge terms in eq. (8). Changes to the Approach III result are of order ¼$\eta$, and therefore insignificant. Adding the dipole-image terms in eq. (7) and/or eq. (8) generates further changes that are insignificant.

Finally, Approach V is a new approximation, introduced because it simplifies the analyses in Sections IV and V. Approach V uses eq. (7) alone, and disregards both the dipole-related terms. This leads to marginally less accurate estimates of apex field and FEF.

TABLE I. To show the strengths ($q$ and $p$) of the sphere charge and sphere dipole, and related potential contributions, for five different mathematical approaches defined in the text. The symbol $\eta$ [$=r/\ell$] denotes the ratio of sphere radius $r$ to sphere elevation $\ell$. $E_M$ denotes conventional macroscopic electrostatic field. All results include terms up of order $\eta^2$. Approaches I to III yield mathematically exact results.

|  | Mathematical approach | | | | |
|---|---|---|---|---|---|
|  | I | II | III | IV | V |
| Strength/$4\pi\varepsilon_0$ | Value / $E_M$ | | | | |
| $q/4\pi\varepsilon_0$ | $r\ell(1+\eta)$ | $r\ell$ | $r\ell(1+\tfrac{1}{2}\eta)$ | $r\ell(1+\tfrac{1}{2}\eta+\tfrac{1}{4}\eta^2)$ | $r\ell(1+3\eta/2+\tfrac{1}{2}\eta^2)$ |
| $p/4\pi\varepsilon_0$ | na | $r^3$ | $r^3$ | $r^3(1-\tfrac{1}{4}\eta-\eta^2/8)$ | na |
| Physical origin | Apex potential contribution / $\ell E_M$ | | | | |
| Sphere-charge | $1+\eta$ | 1 | $1+\tfrac{1}{2}\eta$ | $1+\tfrac{1}{2}\eta+\tfrac{1}{4}\eta^2$ | $1+3\eta/2+\tfrac{1}{2}\eta^2$ |
| Sphere-dipole | na | $\eta$ | $\eta$ | $\eta-\tfrac{1}{4}\eta^2$ | na |
| Macroscopic field | $-1-\eta$ | $-1-\eta$ | $-1-\eta$ | $-1-\eta$ | $-1-\eta$ |
| Image charge | na | na | $-\tfrac{1}{2}\eta$ | $-\tfrac{1}{2}\eta$ | $-\tfrac{1}{2}\eta-\tfrac{1}{2}\eta^2$ |
| TOTAL | 0 | 0 | 0 | 0 | 0 |

TABLE II. To show the field contributions and total apex FEF, for five different mathematical approaches defined in the text. Symbol meanings are as in Table I. Results for approaches I and II are mathematically exact; other results include terms up to $\eta^2$.

|  | Mathematical approach | | | | |
|---|---|---|---|---|---|
|  | I | II | III | IV | V |
| Physical origin | Apex field contribution / $E_M$ | | | | |
| Sphere charge | $(\ell/r)+1$ | $(\ell/r)$ | $(\ell/r)+\tfrac{1}{2}$ | $(\ell/r)+\tfrac{1}{2}+\tfrac{1}{4}\eta+\tfrac{1}{4}\eta^2$ | $(\ell/r)+1.5+\tfrac{1}{2}\eta$ |
| Sphere dipole | na | 2 | 2 | $2-\tfrac{1}{2}\eta+\tfrac{1}{4}\eta^2$ | na |
| Macroscopic field | 1 | 1 | 1 | 1 | 1 |
|  | na | na | $-\tfrac{1}{4}\eta+\eta^2/8$ | $-\tfrac{1}{4}\eta-\eta^2/8$ | $-\tfrac{1}{4}\eta-\eta^2/8$ |
| Total (i.e., $\gamma_{one}=$) | $(\ell/r)+2$ | $(\ell/r)+3$ | $(\ell/r)+3.5$ $-\tfrac{1}{4}\eta+\eta^2/8$ | $(\ell/r)+3.5$ $-\tfrac{1}{2}\eta+\eta^2/8$ | $(\ell/r)+1.5$ $+\tfrac{1}{4}\eta-\eta^2/8$ |

### D. Discussion (single-emitter case)

In the Tables, the best mathematical approximation is IV. The related full FEF-formula is:

$$\gamma^{IV} = \left[\frac{\eta^{-1}+\tfrac{7}{2}-\eta-\tfrac{7}{4}\eta^2-\tfrac{3}{8}\eta^3}{(1-\tfrac{1}{2}\eta-\tfrac{1}{4}\eta^2)(1+\tfrac{1}{2}\eta)}\right], \qquad (10)$$

which on simplification (using MAPLE) generates the approximation

$$\gamma^{IV} = (\ell/r) + \tfrac{7}{2} - \tfrac{1}{2}\eta + \tfrac{1}{8}\eta^2 - \tfrac{3}{16}\eta^3 + O(\eta^5) \,. \tag{11}$$

This confirms the corresponding ZPCL result [their eq. (15)].

Strictly, eqs (7) and (8) do not represent a physically correct analysis of the situation. In reality, the original image-charge $-q^0$ located a distance $\ell$ behind the emitter plate will induce a charge distribution in the sphere, and the effects of this must be cancelled by placing a charge $q^1$ inside the sphere at a position offset from the sphere centre in the direction of the emitter plate[8,30]. This in turn needs an image in the emitter plate, and so on. Similar procedures are needed for the image dipole.

ZPCL give an exact treatment of the problem, and reach a series result [their eq. (6)] that—to third order—reads

$$\gamma^{exact} = (\ell/r) + \tfrac{7}{2} - \tfrac{1}{2}\eta - \tfrac{1}{8}\eta^2 + \tfrac{7}{16}\eta^3 + O(\eta^4) \,. \tag{12}$$

Clearly, the result for Approach IV differs from this by around $\eta^2/4$, which for practical emitters is negligible in comparison with the leading term $(\ell/r)$ $[=\eta^{-1}]$. In practical calculations there is no useful merit in using the exact approach that needs an infinite series of image charges and dipoles.

Aside from the FSEPP model, several more-accurate electrostatic analyses of the "hemisphere on a cylindrical post" geometry exist. These take various detailed forms but coincide in finding that, for moderate to large values of $(\ell/r)$ a better rough approximation for $\gamma_{one}$ is[6]

$$\gamma_{one} \sim 0.7\,(\ell/r) \,. \tag{13}$$

It is clear from inspection of Tables I and II that, when $(\ell/r)$ is adequately large, then this $(\ell/r)$ term dominates, and there is little to choose between the various FSEPP model approximations – especially since the basic model accuracy is only around 30%. Thus, in this paper, we can use whichever is most convenient. This will be Approach V (or Approach I where this is an adequate approximation). It also follows that, in binomial expansions, usually only the leading term is needed.

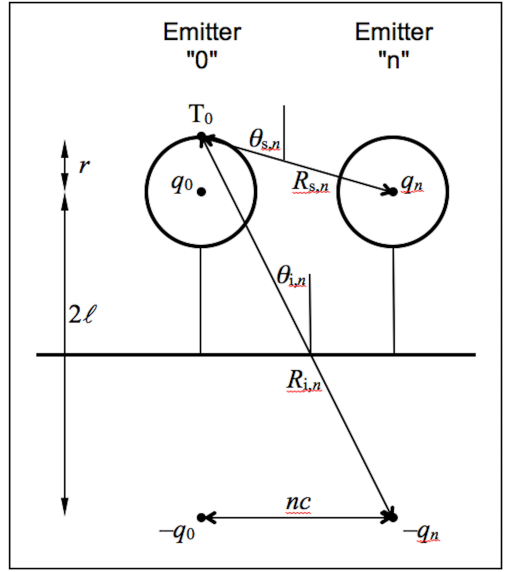

FIG. 2. To illustrate modelling of the two-emitter system, showing relevant geometrical parameters. Each individual emitter is modelled as shown in Fig. 1. The diagram is not to scale.

## IV. THE TWO-EMITTER CASE

### A. Potential contributions and related equations

Consider two identical emitters "0" and "n", separated by a distance written $nc$, and modelled as shown in Fig. 2. At the apex $T_0$ of sphere "0", in addition to contributions discussed above, there will be potential and field contributions due to emitter "n". These can be described as follows.

The distances $R_{s,m}$ and $R_{i,m}$ between $T_0$ and, respectively, the centre of sphere "m" and the centre of the image of sphere "m", are given, for both $m=0$ and $m=n$ ($n\neq 0$) by the formulae

$$R_{s,m} = \{m^2 c^2 + r^2\}^{1/2}, \tag{14}$$

$$R_{i,m} = \{m^2 c^2 + (2\ell + r)^2\}^{1/2}. \tag{15}$$

The sphere and image charges $q_m$ and $-q_m$ for emitter "m" make contributions $\varphi_{0,s,m}$ and $\varphi_{0,i,m}$,

respectively, to the potential at T$_0$, where

$$\varphi_{0,s,m} = q_m/4\pi\varepsilon_0 R_{s,m} , \quad \varphi_{0,i,m} = q_m/4\pi\varepsilon_0 R_{i,m} . \tag{16}$$

It is useful to introduce dimensionless parameters $C_m$ ($m \geq 0$) defined by

$$C_m = r/R_{s,m} - r/R_{i,m} = \{1+(mc/r)^2\}^{-1/2} - \{(mc/r)^2 + (1+2\ell/r)^2\}^{-1/2} . \tag{17a}$$

Specifically, $C_0$ is given by eq. (17b) and can be approximated as shown if $r/\ell \ll 1$:

$$C_0 = 1 - (1+2\ell/r)^{-1} \approx 1 - r/2\ell = 1 - \eta/2 . \tag{17b}$$

For the apex of any sphere "m", $C_0$ can be interpreted physically as the ratio of the potential contribution by the sphere and image charges associated with emitter "m" to the contribution made by the sphere charge for emitter "m" alone. Similarly, $C_m$ ($m \geq 1$) relates to the potential contribution that would be made at T$_0$ by the charges associated with emitter "m" (a horizontal distance $mc$) away, if $q_m$ were equal to $q_0$.

In principle, there are also potential (and field) contributions, at T$_0$, due to sphere *dipole* "m" ($m \geq 1$) and its image. These are relatively small in practical situations and can be neglected here.

In what follows, a different form of the Approach V analysis of the one-emitter case is useful. Denote the sphere-charge by $q_{\text{one}}$, the apex field by $E_{\text{one}}$, and (taking $p=0$) write eq. (7) in the form

$$(q_{\text{one}}/4\pi\varepsilon_0 r)C_0 = (r+\ell)E_M = r(1+\ell/r)E_M . \tag{18}$$

On defining $K$ as below, $E_{\text{one}}$ becomes given by

$$E_{one} = (q_{one}/4\pi\varepsilon_0)[1/r^2 - 1/(2\ell+r)^2] = (q_{one}/4\pi\varepsilon_0 r^2)K ,  \qquad (19a)$$

$$K = 1 - 1/(1+2\ell/r)^2 \approx 1 - (r/2\ell)^2 = 1 - \eta^2/4 \qquad (19b)$$

For a given emitter, $K$ is the parameter that relates the apex field contribution due to both its sphere and image charges to the contribution generated by its sphere charge alone. For practical emitters, $K$ is close to unity and is a weak function of the ratio $\ell/r$. The related apex FEF $\gamma_{one}$ can then be written

$$\gamma_{one} = E_{one}/E_M = (1+\ell/r)K/C_0 \qquad (20)$$

and we may write

$$\gamma_{one}C_0 = (1+\ell/r)K = -(K/E_M r)\varphi_M , \qquad (21)$$

where $\varphi_M$ is the contribution made by the macroscopic field to the potential at the sphere apex. Note that $\gamma_{one}C_0$ is a function only of the ratio $\ell/r$, and that $-(K/E_M r)$ becomes a constant when the values of $r$, $\ell/r$ and $E_M$ are fixed.

## B. Mutual charge-blunting (2-emitter case)

As discussed in Sec. IID, the proximity of two emitters causes mutual charge blunting. If the potential terms due to emitter "n" ($n \geq 1$) are included in an equation equivalent to eq. (7), and all dipole terms are neglected, the result is

$$(q_{two}/4\pi\varepsilon_0 r)(C_0 + C_n) = r(1+\ell/r)E_M \qquad (22)$$

where $q_{two}$ is the common value of $q_0$ and $q_n$. On defining $E_{two}$ [$=K(q_{two}/4\pi\varepsilon_0 r^2)$] and $\gamma_{two}$ [$=E_{two}/E_M$] as the apex field and FEF at $T_0$ resulting from the sphere and image charges for emitter "0" alone, eqs (21) and (22) yield

$$\gamma_{\text{two}}(C_0 + C_n) = \gamma_{\text{one}} C_0 . \tag{23}$$

Physically, what this equation specifies is that in the two-emitter case the potential at $T_0$ must be the same as it was in the one-emitter case, if $r$, $\ell/r$ and $E_M$ are the same in both cases. The change $\Delta\gamma_{\text{two}}$ and fractional change $\delta_{\text{CB}}$, due to mutual charge-blunting (CB), are then found as

$$\delta_{\text{CB}} = \Delta\gamma_{\text{two}}/\gamma_{\text{one}} = (\gamma_{\text{two}} - \gamma_{\text{one}})/\gamma_{\text{one}} = -C_n/(C_0 + C_n) . \tag{24}$$

Within Approach V, eq. (24) is a simple, exact two-emitter result for the effect of mutual charge-blunting, and can be evaluated numerically (see Section VI). For practical emitters, it is always true that $C_0 \gg C_n$, and $C_0 \approx 1$ and hence that

$$\delta_{\text{CB}} \approx -C_n . \tag{25}$$

Since it is always true that $C_n > 0$, mutual charge-blunting produces a *negative* fractional change in apex FEF at all spacings.

There are two regimes where simple explicit approximations exist. For emitters positioned such that $r \ll nc \ll \ell$, called here *closely spaced ("cs") emitters*, eqs. (17) and (25) can be evaluated in an approximation in which the plate-image terms are disregarded. In this case, $C_n \approx r/nc$, and

$$\delta_{\text{CB}}(\text{"cs"}) \approx -(r/nc) = -\eta(\ell/nc) . \tag{26}$$

The condition $r \ll nc$ is important is reaching this result. If spheres are *very* close together, then the approximation of locating the image of sphere charge "n" in sphere "0" to be at the centre of sphere "0" breaks down (for example, see ZPCL, start of their Section III). I can find no clearly formulated numerical analysis of how this breakdown occurs, so it is assumed here that an adequate validity requirement is $nc>4r$, or equivalently $(nc/\ell)>4\eta$.

For two emitters positioned such that $r<<\ell<<nc$, called here *widely spaced ("ws") emitters*, analysis must include the plate-image term for "n". In this case $C_n \approx 2r\ell^2/n^3c^3$ ($n\geq 1$), and

$$\delta_{CB}("ws") \approx -2r\ell^2/n^3c^3 = -2\eta(\ell/nc)^3 ; \tag{27}$$

### C. Neighbor-field effects (2-emitter case)

In terms of $R_{s,n}$, $R_{i,n}$ and the angles $\theta_{s,n}$ and $\theta_{i,n}$ shown in Fig. 2, the sphere and image charges for "n" make contributions $e_{0,s,n}$ and $e_{0,i,n}$ to the field component normal to the sphere surface at $T_0$, where

$$e_{0,s,n} = (q_n/4\pi\varepsilon_0 R_{s,n}^2)\cos\theta_{s,n} = (q_n/4\pi\varepsilon_0)(r/R_{s,n}^3) , \tag{28}$$

$$e_{0,i,n} = -(q_n/4\pi\varepsilon_0 R_{i,n}^2)\cos\theta_{i,n} = -(q_n/4\pi\varepsilon_0)(2\ell+r)/R_{i,n}^3) \tag{29}$$

The sphere and image charges for "n" will also produce, at $T_0$, a field component *parallel* to the surface of sphere "0". This will be cancelled by induced polarization of sphere "0" parallel to the emitter plate, and has no significant effect on the present analysis.

The direct influence of "n" on "0", due to the total neighbor-field (NF) contribution $e_{0,t,n}$ [$=e_{0,s,n}+e_{0,i,n}$] at $T_0$, gives a *further* fractional change $\delta_{NF}$ in the apex FEF of "0" (beyond that due to charge-blunting), with $\delta_{NF}$ given by

$$\delta_{NF} = e_{0,t,n}/E_{one} = (e_{0,t,n}/E_{two})(E_{two}/E_{one}) = (\gamma_{two}/\gamma_{one})(e_{0,t,n}/E_{two}) \tag{30a}$$

$$\delta_{NF} = \{C_0/(C_0+C_n)\}\{e_{0,t,n}/(q_{two}/4\pi\varepsilon_0 r^2)\}. \tag{30b}$$

From equations above, it can be shown[31] that

$$\delta_{NF} = \{C_0/(C_0+C_n)\}K^{-1}[r^3/\{n^2c^2+r^2\}^{3/2} - r^2(2\ell+r)/\{n^2c^2+(2\ell+r)^2\}^{3/2}] \tag{31}$$

When $r<<\ell$, then $K\approx 1$; when $r<<c$, then $C_n<<C_0$ and $\{C_0/(C_0+C_n)\}\approx 1$. When both conditions apply (which is always the case for practical emitters), then eq. (31) reduces to:

$$\delta_{NF} = (r/nc)^3 - 2r^2\ell/\{n^2c^2+(2\ell)^2\}^{3/2}. \tag{32}$$

For closely spaced emitters ($r<<nc<<\ell$), eq. (32) reduces further to

$$\delta_{NF} = (r/nc)^3 - r^2/4\ell^2. \tag{33}$$

Hence, if $(r/nc)^3 > (r/\ell)^2/4$, or equivalently

$$(nc)^3 < 4r\ell^2, \tag{34}$$

then $\delta_{NF}$ will be positive, and *direct FEF-increase* occurs.

A rough estimate of the range of values of ($nc/r$) where this happens is found by using the approximation $\gamma_{one} \sim \ell/r$, and taking 200 as a typical value of $\gamma_{one}$. This yields the rough estimate that direct FEF increase occurs in the range when ($nc/r$) < ~50, or equivalently ($nc/\ell$) < ¼. For most or many practical array geometries, this latter condition would not be met, and consequently the direct effect would be FEF-decrease for all emitter pairs.

In principle, a situation could arise in a regular multi-emitter array whereby, for a given emitter,

nearby emitters provide direct FEF-increase, but emitters further away provide direct FEF-decrease. The net outcome would then need to be established by detailed summations.

When emitters are "sufficiently closely spaced" ("scs") and condition (34) is well satisfied, and significant FEF-increase is occurring, the term in $\ell$ in eq. (33) can be neglected. In this case, $\delta_{NF}$ is given adequately by

$$\delta_{NF}(\text{"scs"}) \approx (r/nc)^3 = +\eta(\ell/nc)^3 \tag{35}$$

In the case of widely separated emitters ($r<<\ell<<nc$), expression (32) reduces to

$$\delta_{NF}(\text{"ws"}) \approx (r-2\ell)r^2/(nc)^3 \approx -2\ell r^2/(nc)^3 \approx -2\eta^2(\ell/nc)^3 \tag{36}$$

In this "widely separated" limit, the effect is always a decrease in apex FEF. Obviously, the size of this neighbor-field-induced FEF-decrease dies away as the separation $nc$ increases.

## D. Discussion (2-emitter case)

In equations above, the emitter spacing has been written as $nc$, as this yields formulas useful for discussing arrays of more than two emitters. When $n$ is put equal to 1, the spacing is $c$, and simplified results for $\delta$ are as in Table III.

TABLE III. Fractional FEF changes ($\delta$) for 2-emitter case. This table records the "leading-term approximations" for the effects and regimes shown.

| Cause | Closely spaced ("cs") ($r<<c<<\ell$) | Widely spaced ("ws") ($r<<\ell<<c$) |
|---|---|---|
| Charge-blunting | $-(r/c) = -\eta(\ell/c)$ | $-2\eta(\ell/c)^3$ |
| Neighbor-field | $+(r/c)^3 = +\eta^3(\ell/c)^3$ [a] | $-2\eta^2(\ell/c)^3$ |

[a] These results also require the condition: $c^3 << 4r\ell^2$.

Clearly, for the FSEPP model, charge-blunting effects are very significantly greater in magnitude than neighbor-field effects in both regimes considered, and (by extension) in all 2-emitter cases of practical interest. For sufficiently closely spaced emitters [$r<<c<<\ell$, and $c^3<4r\ell^2$], the ratio is $(c/r)^2$, which is unlikely to be less than about 25; for widely spaced emitters, the ratio is $1/\eta$ [$\approx \gamma_{\text{one}}$], which might typically be ~200.

This implies that, in FSEPP modelling of electrostatic interactions between identical emitters, it is sufficient (in a first approximation) to consider only charge-blunting effects.

### E. Infinite regular arrays

With an infinite or very large regular array, mathematical arguments as above apply to all emitter pairs. For a given emitter, summations need in principle to be carried out over all other emitters.

As far as the author is aware, all existing FSEPP array models disregard neighbor-field effects, and concentrate analysis on charge-blunting effects. Although this approximation remains to be formally proven for infinite and very large arrays, it is almost certainly valid.

For charge-blunting effects in infinite arrays, ZPCL explore several physical/mathematical approximations, and show that (provided the spheres are not *very* closely spaced) the resulting apex FEF $\gamma_{\text{inf}}$ is given adequately by their "initial approximation" [their eq. (19)]:

$$\gamma_{\text{inf}} = 3 + \frac{1}{\eta + 4\pi(r^2/A_{\text{c}})}, \qquad (37)$$

where $A_{\text{c}}$ is the "emitter footprint", i.e. area per emitter, in the array.

As ZPCL point out, as sphere elevation increases (and hence $\eta \rightarrow 0$), formula (37) tends to the limit

$$\gamma_{\text{inf}} \rightarrow 3 + A_{\text{c}}/4\pi r^2. \qquad (38)$$

Physically, what happens (in this infinite-array case) is that, when all the charge originally on the emitter plate has been transferred to the spheres, then no further charge-sharpening is possible.

In this limit, the main term in eq. (38) has a simple derivation[32]. Without the emitters, the plate's surface charge density is $\varepsilon_0 E_{\text{M}}$, and the charge in the footprint area $A_{\text{c}}$ of a single emitter is $\varepsilon_0 A_{\text{c}} E_{\text{M}}$.

When all this charge is at the sphere centre, it creates an apex field $A_c E_M / 4\pi r^2$ and an apex FEF contribution $A_c / 4\pi r^2$.

Obviously, in the opposite limit of well separated emitters, where $A_c$ becomes large, $\gamma_{inf}$ tends towards $(\eta^{-1}+3)$, which is the formula found earlier for $\gamma_{one}$ via Approach II.

## V. THE EQUISPACED THREE-EMITTER LINEAR ARRAY
### A. Potential and field contributions and related equations

Consider a linear array of three equispaced identical emitters "0", "1" and "2", with spacing $c$. Let the sphere charges after charge-blunting be $q_m$ ($m$= 0,1,2). Since "0" and "2" are equivalent, $q_2 = q_0$; but $q_1 \neq q_0$. Before neighbor-fields are included, let apex fields be $E_m$ [$=Kq_m/4\pi\varepsilon_0 r^2$] and apex FEFs be $\gamma_m$ [$=E_m/E_M$].

In this 3-emitter case, eq. (23) becomes replaced by the two simultaneous equations

$$\gamma_0 (C_0 + C_2) + \gamma_1 C_1 = \gamma_{one} C_0, \tag{39}$$

$$2\gamma_0 C_1 + \gamma_1 C_0 = \gamma_{one} C_0. \tag{40}$$

Equations (39) and (40) in effect specify that the potentials at the apexes of (respectively) spheres "0" and "1" must be the same as the potential at $T_0$ in the one-emitter case.

Solution by subtraction and re-arrangement yields[31] a formula for the *exposure ratio* $\Xi_{CB}$ (of "0", relative to "1", due to differential charge blunting):

$$\Xi_{CB} \equiv \gamma_0 / \gamma_1 = (C_0 - C_1) / (C_0 - 2C_1 + C_2) \tag{41}$$

Since $C_0 \gg C_1$ and $C_0 \gg C_2$, and also $C_0 \approx 1$, a binomial expansion of eq. (41) yields the approximation

$$(\Xi_{CB} - 1) \approx (C_1 - C_2) / C_0 \approx C_1 - C_2 \tag{42}$$

Since $C_1 > C_2$, this result shows that $\gamma_0/\gamma_1 > 1$, i.e., that differential charge-blunting makes the outer emitters have a higher apex-FEF value than the central emitter, at all spacings.

Equations (39) and (40) can be used to find expressions for $\gamma_0$ and $\gamma_1$ in terms of $\gamma_{one}$. For emitter "m", the change $\Delta\gamma_m$ and fractional change $(\delta_m)_{CB}$ in apex FEF, due to charge-blunting (including differential charge-blunting) are then found from

$$(\delta_m)_{CB} \equiv \Delta\gamma_m / \gamma_{one} \equiv (\gamma_m - \gamma_{one})/\gamma_{one} \ . \tag{43}$$

Algebraic analysis[31] yields the approximations

$$(\delta_0)_{CB} \approx -(C_1 + C_2) \ , \tag{44}$$

$$(\delta_1)_{CB} \approx -2C_1 \ . \tag{45}$$

The exposure-ratio component $\Xi_{CB}$ can also be expanded formally as:

$$\Xi_{CB} = \frac{1+(\delta_0)_{CB}}{1+(\delta_1)_{CB}} \approx 1+(\delta_0)_{CB} - (\delta_1)_{CB} \ . \tag{46}$$

By analogy, the total exposure ratio $\Xi_{tot}$ (i.e., total FEF at site "0" divided by total FEF at site "1") can be written (when both are divided by $\gamma_{one}$):

$$\Xi_{tot} = \frac{1+(\delta_0)_{CB}+(\delta_0)_{NF}}{1+(\delta_1)_{CB}+(\delta_1)_{NF}} \approx 1+(\delta_0)_{CB}-(\delta_1)_{CB}+(\delta_0)_{NF}-(\delta_1)_{NF} \ , \tag{47}$$

and the neighbor-field contribution $\Xi_{NF}$ can be written

$$\Xi_{\mathrm{NF}} \approx \frac{1+(\delta_0)_{\mathrm{NF}}}{1+(\delta_1)_{\mathrm{NF}}} \approx 1+(\delta_0)_{\mathrm{NF}} - (\delta_1)_{\mathrm{NF}}. \tag{48}$$

Further, to an adequate approximation

$$\Xi_{\mathrm{tot}} = \Xi_{\mathrm{CB}} \times \Xi_{\mathrm{NF}}. \tag{49}$$

The fractional changes in apex FEF, due to neighbor-field effects, are given formally by

$$(\delta_2)_{\mathrm{NF}} = (\delta_0)_{\mathrm{NF}} = (e_{0,\mathrm{t},1} + e_{0,\mathrm{t},2})/E_{\mathrm{one}}, \tag{50}$$

$$(\delta_1)_{\mathrm{NF}} = 2e_{1,\mathrm{t},0}/E_{\mathrm{one}}. \tag{51}$$

## B. Results and discussion (3-emitter case)

In the 3-emitter case, the most interesting parameters are $\delta_1$ and $(\Xi-1)$. Detailed algebraic analysis[31] yields the results shown in Tables IV and V.

TABLE IV. Fractional FEF changes ($\delta_1$) for the central emitter in the 3-emitter situation. This table records the "leading-term approximations" for the effects and regimes shown.

| Cause | Closely spaced ("cs") $(r<<c<<\ell)$ | Widely spaced ("ws") $(r<<\ell<<c)$ |
|---|---|---|
| Charge-blunting | $-2(r/c) = -2\eta\,(\ell/c)$ | $-4\eta\,(\ell/c)^3$ |
| Neighbor-field | $+2(r/c)^3 = +2\eta^3(\ell/c)^3$ [a] | $-4\eta^2(\ell/c)^3$ |

[a] These results also require the condition: $c^3 << 4r\ell^2$.

TABLE V. Values for the quantity $(\Xi-1)$ in 3-emitter situation. This table records the "leading-term approximations" for the effects and regimes shown.

| Cause | Closely spaced ("cs") $(r<<c<<\ell)$ | Widely spaced ("ws") $(r<<\ell<<c)$ |
|---|---|---|
| Charge-blunting | $+\frac{1}{2}(r/c) = +\frac{1}{2}\eta(\ell/c)$ | $+(7/4)\eta(\ell/c)^3$ |
| Neighbor-field | $-(7/8)(r/c)^3 = -(7/8)\eta^3(\ell/c)^3$ [a] | $+(7/4)\eta^2(\ell/c)^3$ |

[a] These results also require the condition: $c^3 << 2r\ell^2$.

As might be expected from the general linearity of basic electrostatics, comparisons of Tables III and IV show that, for the central emitter in 3-emitter geometry, the FEF changes are, in each case, simply twice the related FEF change in 2-emitter geometry. By extension, this can be generalised to mean that, for the "central emitter" in FSEPP-type models of emitter arrays, charge-blunting effects will always be very significantly greater than neighbour-field effects.

With differential charge-blunting, the positive ($\Xi$–1) values show that the apex FEF for the "more exposed" outer emitters ("0" and "2") is higher at all spacings than the apex FEF for the central emitter. With differential neighbor-field effects, this is true only for well separated emitters; by contrast, for sufficiently closely separated emitters the effect is reversed, and the apex FEF for the central emitter is increased more than the apex FEF for the outer emitters.

The differential effects due to charge-blunting are significantly larger than the differential effects due to neighbor-fields; thus, for practical emitters, it will always be the case that the apex FEF for the central emitter will be less than those for the outer emitters. By extension, for larger linear arrays one expects the apex FEF values to be higher near the ends of the array. With two-dimensional arrays one expects the apex-FEF values to be highest at the corners of the array, because the corner emitters have fewer neighbors and are most "exposed". These FSEPP-model implications of differential charge-blunting are, of course, consistent with Laplace-type numerical analyses of emitter arrays.[16-18]

## VI. GENERAL DISCUSSION

### A. Summary

The prime technological motivations for work on the electrostatics of field emitter arrays/clusters are to understand how to design array geometry so that: (a) the average (or "macroscopic") current density for a given macroscopic field $E_M$ is a maximum (subject to any fabrication constraints); and (b) the current per emitter is the same for all emitters. A good understanding of the physical electrostatics of arrays/clusters, and associated mathematics, is a desirable preliminary. This paper aimed to improve understanding. Using the "floating sphere at emitter-plane potential" model, it has looked in turn at the single emitter, the 2-emitter system and the linear 3-emitter array.

For a single emitter, various approaches were noted. A simple approximation that considers only sphere charges and (for widely separated emitters) image charges is adequate for the present work.

In the 2-emitter system, two modification effects on apex-FEF values were examined, namely charge-distribution effects and neighbor-field effects. With the former, for fixed ($\ell/r$) the effect is always mutual charge-blunting, and hence reduction of the apex-FEF $\gamma_{two}$ relative to the 1-emitter value $\gamma_{one}$. The blunting gets more pronounced as spacing decreases, as has been found in FSEPP modelling of large emitter arrays (e.g., Ref. 33). Neighbor-field effects cause apex-FEF increase at sufficiently small emitter spacing $c$, but apex-field decrease at larger spacings. For practical emitters, charge-blunting effects are always much greater in magnitude than neighbor-field effects. Similar considerations can be assumed to apply to infinite regular arrays.

With a FSEPP model of a 3-emitter array, differential effects occur. Charge-blunting effects always *decrease* the central "interior" apex FEF more than the outer ("edge") FEFs. Neighbor-field effects *increase* the interior apex FEF more than the outer apex FEFs when the spacing is sufficiently small, but *decrease* the interior apex FEF more than the outer apex FEFs when the spacing is sufficiently large. The differential effects due to charge-blunting are always much greater in magnitude than the differential effects due to neighbor-fields.

By extension, the above results mean that, in FSEPP models, neighbor-field effects can always be disregarded in practice. Hence, for larger linear arrays and for two-dimensional arrays, the evaluation of fractional FEF changes should be a relatively straightforward matter, using charge-blunting type equations analogous to eqs (39) and (40). For example, for a linear array of five equally spaced identical emitters {"0" to "4"} the set of equations would be

$$\gamma_0(C_0 + C_4) + \gamma_1(C_1 + C_3) + \gamma_2 C_2 = \gamma_{one} C_0 , \quad (52a)$$

$$\gamma_0(C_1 + C_3) + \gamma_1(C_0 + C_2) + \gamma_2 C_1 = \gamma_{one} C_0 , \quad (52b)$$

$$2\gamma_0 C_2 + 2\gamma_1 C_1 + \gamma_2 C_0 = \gamma_{one} C_0 . \quad (52c)$$

These can be solved by standard matrix-algebra methods.

The physical effect called here "differential charge blunting" is, in principle, an extremely well known effect. It is the primary reason why fields are highest at the sharpest points of three-dimensional conductors. It has been described[34] in classical electrostatics, and a three-dimensional atomic-level equivalent has recently been used to discuss field evaporation from a field ion emitter.[35] The present paper has demonstrated that the same basic electrostatic principle (derived ultimately from electron thermodynamics) applies to field emitter arrays.

**B. Line-charge models**

As noted earlier, FSEPP models do not provide quantitatively accurate FEF values. Two methods of potentially higher accuracy are (a) use of better analytical emitter models, for example line-charge models (LCMs), and (b) analysis by numerical solution of Laplace's equation. LCMs for field emitters were introduced some years ago (e.g., Ref. 36), and have recently been used by Harris, Shiffler, Jensen and colleagues (e.g., Refs 15,19).

With real field emitters the charge spreads down the emitter sides: LCMs can take this into account. The spreading tends both to weaken charge-blunting effects and strengthen neighbor-field effects. However, it provisionally appears that recent LCM-model applications may, on the face of it, overweight neighbor-field effects as compared with charge-blunting effects. By adjustment of the LCM, by comparison with Laplace-solver numerical results, it is possible to get results of correct magnitude. However, LCM models as used, for example, in Ref. 15 (see their Fig. 4), do not have the behavior qualitatively expected from the FSEPP model as emitter spacing decreases. In Fig. 4, the apex FEF first decreases and then increases again, as spacing decreases. This is in marked contrast to the behavior expected from charge blunting, which is steady (and increasingly rapid) decrease in apex FEF as spacing decreases. This steady decrease is apparent in earlier equations, and is also found in FSEPP models of large emitter arrays (e.g., Fig. 4 in Ref. 33) and—*crucially*—in some existing Laplace-based calculations (e.g., Fig. 9a in Ref. 37).

However, in some recent new Laplace-based calculations on cylindrical-post emitters, a small increase in apex FEF has been found (for the 2-emitter system) at very small [<0.1] values of ($c/\ell$) (Dall'Agnol and de Assis, private communication, 2016). This, of course, is qualitatively consistent with the LCM results in Ref. 15. This effect needs to be investigated further.

With the steady increase in computer power and increasing availability of automatic meshing techniques, it is arguable that Laplace-equation-based numerical techniques are (or will become) a superior method for field emitter array analysis. Such techniques ought to be more accurate than analytical models, but will be most useful if numerical results for apex FEFs can be well fitted with a suitable analytical expression.

**C. Fitting of Laplace-based numerical results**

Existing fitting formulas[7,38,39] for Laplace-based numerical results are phenomenological. In particular, the Bonard et al. fitting formula[38], as reformulated by Jo et al.[39] gives (in present notation) the apex FEF $\gamma$ for emitters in a small array as

$$\gamma = \gamma_{\mathrm{one}}[1 - \exp\{-2.3172(c/\ell)\}] . \tag{53}$$

The related fractional reduction $(-\delta)$ in apex FEF can be written

$$\ln(-\delta) = -2.3172(c/\ell) . \tag{54}$$

An alternative might be to use a fitting formula derived by analogy with the 2-emitter analysis above. For two widely spaced emitters, eq. (27) (with $n=1$) adequately gives the fractional reduction $-\delta_{\mathrm{two}}$("ws") in apex FEF as $2\eta\,(\ell/c)^3$. On a ln-ln plot, made against $\ln(c/\ell)$, the large-$c$ form would be the straight line

$$\ln(-\delta_{\mathrm{two}}) \approx \ln(2\eta) - 3\ln(c/\ell) . \tag{55}$$

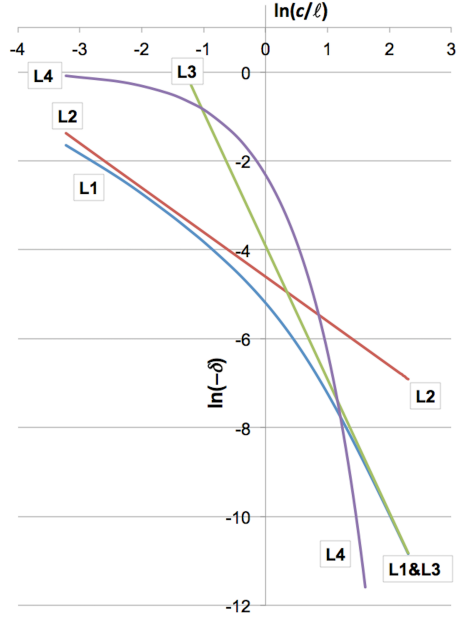

FIG. 3 (Color online.) Logarithmic diagram, showing how the fractional reduction ($-\delta$) in apex FEF, for the two-emitter case, depends on the ratio ($c/\ell$) of emitter spacing $c$ to sphere elevation $\ell$. An illustrative value of $\ell/r =100$ has been assumed, where $r$ is sphere radius. Line L1 is the precise result from the FSEPP model (ignoring neighbor-field effects); L2 is the simple approximation for behaviour at sufficiently low ($c/\ell$) values; L3 is the limiting behaviour at large ($c/\ell$) values; L4 is the relationship (based on numerical solution of Laplace's equation) derived from Refs. 38 and 39. Results are cut-off at a low-spacing validity limit equivalent to ($c/r$)=4. Note the significant difference between the behaviors of lines L1 and L4.

Similarly, using eq. (26), the small-$c$ form (down to some validity limit) would be the straight line

$$\ln(-\delta_{two}) \approx \ln(\eta) - \ln(c/\ell) . \qquad (53)$$

For the illustrative value ($\ell/r$) =100, these two limiting lines are shown in Fig. 3, together with a curve derived by exact evaluation of eq. (24). These two lines cross when ($c/\ell$)=√2.

However, the *forms* of the fall-off of $-\delta_{two}$, as given via the FSEPP model [eq. (51)], and of $-\delta$, as given via the conventional fitting formula [eq. (50)], are mathematically different, as shown in Fig. 3. Both yield increasingly slow positive change in apex FEF as spacing $c$ increases, but the change happens differently. It should be possible to find the better fitting option by systematic Laplace-type 2-emitter simulations, but insufficient results are currently available.

**D. General conclusions**

The physical electrostatics of field emitter arrays is primarily determined by the effect called here "charge-blunting". However, in the long term, the best way forwards for analysing array electrostatics may be to concentrate on developing high-quality Laplace-based numerical calculations, and reliable fitting equations of known accuracy.

Equations derived from FSEPP models may provide a guide for formulating fitting equations, but an immediate task is to investigate this, as indicated above. It may be that a single fitting equation can be found that covers all regimes of practical interest with sufficient accuracy; or it may be better to use different equations in different parameter regimes.

Another useful step would be to develop precise mathematical analyses of FSEPP models of infinite and very large regular arrays, both to serve as accuracy/validity checks on existing analyses (some of which are approximate), and, perhaps, to provide an alternative guide for formulating fitting equations.

Until such time as good fitting equations to Laplace-based solutions exist, there may be merit in using the FSEPP method to model the electrostatics of small emitter clusters, to provide "quick and easy" rough estimates of the differential effects to be expected in different cluster sizes and configurations. It may also be possible to use a modified version of the method to give useful indicative results in situations where emitters are not geometrically identical and/or are spaced irregularly.

**ACKNOWLEDGMENTS**

An important stimulus for the development of the ideas in this paper was provided by receipt of a preprint (relating to line-charge models of small emitter arrays) written by Harris, Jensen, Tang and Shiffler[40]. The author also thanks Dr J. H. B. Deane for making algebraic checks using the MAPLE™ computer algebra package, and the University of Surrey for provision of facilities.

**Physical electrostatics of small field emitter arrays/clusters**

**Richard G. Forbes**

**SUPPLEMENTARY MATERIAL**

## Details of algebraic analysis

For transparency, and to enable checking, this supplementary material gives details of the algebraic analysis leading to indicated results in the paper. All parameters have the same meaning as in the main text. Equation numbers not prefixed by "S" refer to equations in the main text.

### Parameters and contributions

*The dimensionless parameters $C_m$*

From eqs (14) and (15), the definitions of the distances $R_{s,m}$ and $R_{i,m}$ were ($m \geq 0$)

$$R_{s,m} = \{m^2c^2 + r^2\}^{1/2}, \qquad (14) \ (S1)$$

$$R_{i,m} = \{m^2c^2 + (2\ell+r)^2\}^{1/2}. \qquad (15) \ (S2)$$

For practical emitters $r \ll \ell$, hence it is adequate to assume

$$R_{i,m} \approx \{m^2c^2 + (2\ell)^2\}^{1/2}. \qquad (S3)$$

As in main text, define dimensionless parameters $C_m$ ($m \geq 0$)

$$C_m = r/R_{s,m} - r/R_{i,m} \quad (m \geq 0). \qquad (17) \ (S4)$$

Using (S1) and (S3), an adequate formula for $C_m$ (for practical emitters) is:

$$C_m \approx r/\{m^2c^2 + r^2\}^{1/2} - r/\{(mc)^2 + (2\ell)^2\}^{1/2} \quad (m \geq 0). \qquad (S5)$$

It is usually satisfactory to assume that $r \ll c$, in which case we get

$$C_0 \approx 1, \qquad (S6)$$

$$C_n \approx r/nc - r/\{(nc)^2 + (2\ell)^2\}^{1/2} \quad (n \geq 1). \qquad (S7)$$

*Field contributions $e_{0,t,n}$*

The field symbol $e_{0,t,n}$ denotes the total ("t") field contribution, at the apex $T_0$ of emitter "0", due to the sphere and image charges associated with emitter "n" ($n \geq 1$). From eqs. (28) and (29), $e_{0,t,n}$ is given by

$$e_{0,t,n} = e_{0,s,n} + e_{0,i,n} = (q_n/4\pi\varepsilon_0) \, r/R_{s,n}^3 - (q_n/4\pi\varepsilon_0)(2\ell+r)/R_{i,n}^3, \qquad (S8)$$

$$e_{0,t,n} = (q_n/4\pi\varepsilon_0 r^2)\,[r^3/R_{s,n}^3 - r^2(2\ell+r)/R_{i,n}^3]. \qquad (S9)$$

Using (S1) and (S2), the exact formula becomes

$$e_{0,t,n} = (q_n/4\pi\varepsilon_0 r^2)\,[r^3/\{n^2c^2 + r^2\}^{3/2} - r^2(2\ell+r)/\{n^2c^2 + (2\ell+r)^2\}^{3/2}]. \qquad (S10)$$

In the case of the 2-emitter system, $q_n$ is $q_{\text{two}}$, and it follows that:

$$\{e_{0,t,n}/(q_{\text{two}}/4\pi\varepsilon_0 r^2)\} = [r^3/\{n^2c^2 + r^2\}^{3/2} - r^2(2\ell+r)/\{n^2c^2 + (2\ell+r)^2\}^{3/2}]. \qquad (S11)$$

This result is used in deriving eq. (31) in the main text.

More generally, using (S10), for practical emitters $r \ll \ell$, hence it is adequate to assume

$$e_{0,t,n} \approx (q_n/4\pi\varepsilon_0 r^2) [r^3/\{n^2c^2 + r^2\}^{3/2} - 2r^2\ell/\{n^2c^2 + (2\ell)^2\}^{3/2}]. \quad (S12)$$

If it is also assumed that $r \ll c$, this becomes (note $n \geq 1$)

$$e_{0,t,n} \approx (q_n/4\pi\varepsilon_0 r^2) [(r/nc)^3 - 2r^2\ell/\{n^2c^2 + (2\ell)^2\}^{3/2}]. \quad (S13)$$

## Analysis of the 3-emitter case

### *Solution of the FEF simultaneous equations*

For the case of the linear 3-emitter array, the FEF simultaneous equations are eqs (39) and (40):

$$\gamma_0(C_0+C_2) + \gamma_1 C_1 = \gamma_{one} C_0, \quad (39)\,(S14)$$

$$2\gamma_0 C_1 + \gamma_1 C_0 = \gamma_{one} C_0. \quad (40)\,(S15)$$

Hence:
$$\gamma_0(C_0-2C_1+C_2) + \gamma_1(C_1-C_0) = 0, \quad (S16)$$

$$\gamma_0(C_0-2C_1+C_2) = \gamma_1(C_0-C_1), \quad (S17)$$

$$\gamma_0 = \gamma_1 \frac{C_0 - C_1}{C_0 - 2C_1 + C_2} \quad (S18)$$

This provides eq. (41) in the paper. Further, because $C_0 \gg C_1$ and $C_0 \gg C_2$

$$\Xi_{CB} = \gamma_0/\gamma_1 = \frac{1 - C_1/C_0}{1 - (2C_1 - C_2)/C_0} \approx (1 - C_1/C_0)\{1 + (2C_1 - C_2)/C_0\} = 1 + C_1/C_0 - C_2/C_0, \quad (S19)$$

and, because $C_0 \approx 1$

$$(\Xi_{CB} - 1) \approx C_1/C_0 - C_2/C_0 \approx C_1 - C_2. \quad (S20)$$

This provides eq. (42) in the paper.

Substituting eq. (S18) into eq. (S15) yields

$$\gamma_1 \left[ \frac{2C_1(C_0 - C_1)}{C_0 - 2C_1 + C_2} + C_0 \right] = \gamma_{one} C_0 \tag{S19}$$

$$\gamma_1 \left[ \frac{2C_0 C_1 - 2C_1^2 + C_0^2 - 2C_0 C_1 + C_0 C_2}{C_0 - 2C_1 + C_2} \right] = \gamma_{one} C_0 \tag{S20}$$

$$\gamma_1 \left[ \frac{C_0^2 - 2C_1^2 + C_0 C_2}{C_0 - 2C_1 + C_2} \right] = \gamma_{one} C_0 \tag{S21}$$

$$\gamma_1 = \gamma_{one} \left[ \frac{C_0(C_0 - 2C_1 + C_2)}{C_0^2 - 2C_1^2 + C_0 C_2} \right] \tag{S22}$$

$$\gamma_1 = \gamma_{one} \left[ \frac{(C_0^2 - 2C_0 C_1 + C_0 C_2)}{C_0^2 - 2C_1^2 + C_0 C_2} \right] \tag{S23}$$

$$\Delta\gamma_1 = \gamma_{one} \left[ \frac{(C_0^2 - 2C_0 C_1 + C_0 C_2)}{C_0^2 - 2C_1^2 + C_0 C_2} - 1 \right] = \gamma_{one} \left[ \frac{(C_0^2 - 2C_0 C_1 + C_0 C_2) - C_0^2 + 2C_1^2 - C_0 C_2}{C_0^2 - 2C_1^2 + C_0 C_2} \right] \tag{S24}$$

$$(\delta_1)_{indir} = \Delta\gamma_1 / \gamma_{one} = \left[ \frac{-2C_1 C_0 + 2C_1^2)}{C_0^2 + 2C_1^2 + C_0 C_2} \right] = \left[ \frac{-2C_1(C_0 - C_1)}{C_0^2 + 2C_1^2 + C_0 C_2} \right] \tag{S25}$$

Because $C_0 \gg C_1$ and $C_0 \gg C_2$, and also $C_0 \approx 1$, this reduces to

$$(\delta_1)_{indir} \approx -2C_1. \tag{S26}$$

This provides eq. (45) in the paper, and is twice the 2-emitter result.

From eqs. (S18) and (S22):

$$\gamma_0 = \frac{C_0 - C_1}{C_0 - 2C_1 + C_2} \gamma_1 \tag{S18}$$

$$\gamma_1 = \gamma_{one} \left[ \frac{C_0(C_0 - 2C_1 + C_2)}{C_0^2 - 2C_1^2 + C_0 C_2} \right] \tag{S22}$$

$$\gamma_0 = \frac{C_0 - C_1}{C_0 - 2C_1 + C_2} \gamma_{one} \left[ \frac{C_0(C_0 - 2C_1 + C_2)}{C_0^2 - 2C_1^2 + C_0 C_2} \right] \tag{S27}$$

$$\gamma_0 = \gamma_{one} \left[ \frac{C_0(C_0 - C_1)}{C_0^2 - 2C_1^2 + C_0 C_2} \right] = \gamma_{one} \left[ \frac{C_0^2 - C_0 C_1}{C_0^2 - 2C_1^2 + C_0 C_2} \right] \tag{S28}$$

$$(\delta_0)_{CB} = \left[\frac{C_0^2 - C_0C_1}{C_0^2 - 2C_1^2 + C_0C_2} - 1\right] = \left[\frac{C_0^2 - C_0C_1 - C_0^2 + 2C_1^2 - C_0C_2}{C_0^2 - 2C_1^2 + C_0C_2}\right] = \frac{-C_0C_1 + 2C_1^2 - C_0C_2}{C_0^2 - 2C_1^2 + C_0C_2} \quad (S29)$$

Because $C_0 \gg C_1$ and $C_0 \gg C_2$

$$(\delta_0)_{CB} \approx \frac{-C_0(C_1+C_2)}{C_0^2 - 2C_1^2 + C_0C_2} \approx \frac{-(C_1+C_2)}{C_0} \quad . \tag{S30}$$

And, because $C_0 \approx 1$, we get

$$(\delta_0)_{CB} \approx -(C_1+C_2) \quad . \tag{S31}$$

This provides eq. (44) in the paper.

### *3-emitter charge-blunting effects*

From eqs. (S6) and (S7) above

$$C_0 \approx 1, \tag{S6}$$

$$C_n \approx r/nc - r/\{(nc)^2 + (2\ell)^2\}^{1/2} \quad (n \geq 1). \tag{S7}$$

There are now the same two "limits" as occur in the 2-emitter case

**When $c \ll \ell$,** we get: $\quad C_n \approx r/nc - r/2\ell \approx r/nc$ . (S14)

Hence: $\quad C_1 \approx r/c = \eta\,(\ell/c),$ (S32)

$$C_2 \approx \tfrac{1}{2}\,r/c = \tfrac{1}{2}\,\eta\,(\ell/c), \tag{S33}$$

$$C_1 - C_2 \approx \tfrac{1}{2}\,\eta(\ell/c) . \tag{S34}$$

Hence, using eq. (44): $\quad (\delta_0)_{CB} \approx -(C_1+C_2) \approx -(3/2)\,\eta(\ell/c) .$ (S35)

Hence, using eq. (45): $\quad (\delta_1)_{CB} \approx -2C_1 \approx -2\eta\,(\ell/c) .$ (S36)

Hence, using eq. (42): $\quad (\Xi_{CB}-1) \approx C_1 - C_2 \approx +\tfrac{1}{2}\,\eta(\ell/c) .$ (S37)

**When $\ell \ll c$**, we get $\quad C_n \approx r/nc - (r/nc)/\{1 + (2\ell/nc)^2\}^{1/2} ,$ (S38)

$$C_n \approx r/nc - (r/nc)\{1 - \tfrac{1}{2}(2\ell/nc)^2\} , \tag{S39}$$

$$C_n \approx 2(r/nc)(\ell/nc)^2\} = 2r\ell^2/(nc)^3 . \tag{S40}$$

Hence: $\quad C_1 \approx 2r\ell^2/c^3 = 2\eta(\ell/c)^3,$ (S41)

$$C_2 \approx 2r\ell^2/8c^3 = (\eta/4)(\ell/c)^3, \tag{S42}$$

$$C_1 - C_2 \approx (7/4)\,\eta(\ell/c)^3 \tag{S43}$$

Hence, using eq. (44):
$$(\delta_0)_{CB} \approx -(C_1+C_2) \approx -(9/4)\,(\ell/c)^3 . \tag{S44}$$

Hence, using eq. (45):
$$(\delta_1)_{CB} \approx -2C_1 \approx -4\eta(\ell/c)^3 . \tag{S45}$$

Hence, from eq. (42):
$$(\Xi_{CB}-1) \approx C_1-C_2 \approx +(7/4)\eta(\ell/c)^3 . \tag{S46}$$

The above four results provide the top lines in Tables IV and V.

*3-emitter neighbour-field effects*

It is convenient to start from eq. (S9) above:

$$e_{0,t,n} = (q_n/4\pi\varepsilon_0 r^2)\,[r^3/R_{s,n}^3 - r^2(2\ell+r)/R_{i,n}^3] . \tag{S9}\,(S47)$$

For practical emitters, $r \ll \ell$, so this reduces to

$$e_{0,t,n} = (q_n/4\pi\varepsilon_0 r^2)[r^3/R_{s,n}^3 - 2r^2\ell/R_{i,n}^3] . \tag{S48}$$

Explicitly:
$$e_{0,t,1} = (q_1/4\pi\varepsilon_0 r^2))[r^3/R_{s,1}^3 - 2r^2\ell/R_{i,1}^3] , \tag{S49}$$

$$e_{0,t,2} = (q_0/4\pi\varepsilon_0 r^2)[r^3 R_{s,2}^3 - 2r^2\ell/R_{i,2}^3] , \tag{S50}$$

$$e_{1,t,0} = (q_0/4\pi\varepsilon_0 r^2)[r^3/R_{s,1}^3 - 2r^2\ell/R_{i,1}^3] . \tag{S51}$$

From eqs. (46) and (47), the *further* fractional changes in apex FEF, due to neighbor-field effects, are

$$(\delta_2)_{NF} = (\delta_0)_{NF} = (e_{0,t,1}+e_{0,t,2})/E_{one} , \tag{46}\,(S52)$$

$$(\delta_1)_{NF} = 2e_{1,t,0}/E_{one} . \tag{47}\,(S53)$$

These are now dealt with in turn, with $(\delta_1)_{NF}$ first because it is both simpler and more important. In this case:

$$(\delta_1)_{NF} = (1/4\pi\varepsilon_0 r^2 E_{one})\,2q_0\,\{r^3/R_{s,1}^3 - 2r^2\ell/R_{i,1}^3\} \tag{S54}$$

$$(\delta_1)_{NF} = (q_0/4\pi\varepsilon_0 r^2 E_0)\,(E_0/E_{one})\,2\{r^3/R_{s,1}^3 - 2r^2\ell/R_{i,1}^3\} \tag{S55}$$

Since $(\gamma_0/\gamma_{one}) = (E_0/E_{one})$ (by definition of what $E_{one}$ and $\gamma_{one}$ mean), eq. (S55) becomes

$$(\delta_1)_{NF} = (\gamma_1/\gamma_{one})(q_0/4\pi\varepsilon_0 r^2 E_0)\,2\{r^3/R_{s,1}^3 - 2r^2\ell/R_{i,1}^3\} \tag{S56}$$

Now, by analogy with eq. (19) in the paper, $E_0 = (q_0/4\pi\varepsilon_0 r^2)K$, so $(q_0/4\pi\varepsilon_0 r^2 E_0) = K^{-1}$, and

$$(\delta_1)_{NF} = (\gamma_0/\gamma_{one})K^{-1}\,2\{r^3/R_{s,1}^3 - 2r^2\ell/R_{i,1}^3\} . \tag{S57}$$

Now, for practical emitters $K \approx 1$. Let us also initially work in an approximation in which $(\gamma_0/\gamma_{one})$ and $(\gamma_1/\gamma_{one})$ are both set equal to 1. Hence we obtain:

$$(\delta_1)_{NF} \approx 2\{r^3/R_{s,1}{}^3 - 2r^2\ell/R_{i,1}{}^3\}, \tag{S58}$$

and, on substituting for the "R"s, using eqs (S1) and (S3):

$$(\delta_1)_{NF} \approx 2[r^3/\{c^2+r^2\}^{3/2} - 2r^2\ell/\{c^2+(2\ell)^2\}^{3/2}]. \tag{S59}$$

When $r \ll c$, this reduces to:

$$(\delta_1)_{NF} \approx 2[r^3/c^3 - 2r^2\ell/\{c^2+(2\ell)^2\}^{3/2}] \tag{S60}$$

For small separations ($r \ll c \ll \ell$), this reduces to

$$(\delta_1)_{NF} \approx 2[r^3/c^3 - (1/4)r^2/\ell^2] \tag{S61}$$

$$(\delta_1)_{NF} \approx 2(r/c)^3 - (1/2)(r/\ell)^2. \tag{S62}$$

**For sufficiently small separations [($r \ll c \ll \ell$) and ($c^3 \ll 4r\ell^2$) and ($c > 4r$)]**

$$(\delta_1)_{NF} \approx +2(r/c)^3 = +(9/8)\eta^3 (\ell/c)^3 \tag{S63}$$

**For large separations ($r \ll \ell \ll c$),** eq. (S60) reduces to:

$$(\delta_1)_{NF} \approx 2[r^3/c^3 - 2r^2\ell/c^3] \tag{S64}$$

$$(\delta_1)_{NF} \approx 2(r^2/c^3)[r - 2\ell] \tag{S65}$$

Hence (because $r \ll \ell$)

$$(\delta_1)_{NF} \approx -4\, r^2\ell/c^{3+} = -4\eta^2(\ell/c)^3 \tag{S66}$$

These results provide the bottom line in Table IV.

Because the neighbour-field results are less than the charge-blunting results by a factor of order $\eta^{-1}$ (typically 100 or more), the approximation made in going from eq. (S57) to eq. (S58) is acceptable in a qualitative argument. It would in principle be possible to get a more precise result by iteration, but this does not seem necessary in the context of the aims of this paper.

Expressions for $(\delta_0)_{NF}$ can be obtained in a similar manner. From eqs (S52), (S49) and (S50)

$$(\delta_0)_{NF} = (1/4\pi\varepsilon_0 r^2 E_{one})[q_1\{r^3/R_{s,1}{}^3 - 2r^2\ell/R_{i,1}{}^3\} + q_0\{r^3 R_{s,2}{}^3 - 2r^2\ell/R_{i,2}{}^3\}] \tag{S67}$$

$$(\delta_0)_{NF} = (q_1/4\pi\varepsilon_0 r^2 E_1)(E_1/E_{one})\{r^3/R_{s,1}{}^3 - 2r^2\ell/R_{i,1}{}^3\} + (q_0/4\pi\varepsilon_0 r^2 E_0)(E_0/E_{one})\{r^3 R_{s,2}{}^3 - 2r^2\ell/R_{i,2}{}^3\} \tag{S68}$$

Since $(\gamma_1/\gamma_{one}) = (E_1/E_{one})$ (by definition), and similarly for "0", eq. (S68) becomes

$$(\delta_0)_{NF} = (\gamma_1/\gamma_{one})(q_1/4\pi\varepsilon\, r^2 E_1)\, r^3/R_{s,1}{}^3 - 2r^2\ell/R_{i,1}{}^3\} + (\gamma_0/\gamma_{one})(q_0/4\pi\varepsilon_0 r^2 E_0)\{r^3 R_{s,2}{}^3 - 2r^2\ell/R_{i,2}{}^3\} \tag{S69}$$

Now: $E_1 = (q_1/4\pi\varepsilon_0 r^2)K$, so $(q_1/4\pi\varepsilon\, r^2 E_1) = K^{-1}$, and similarly for "0", so

$$(\delta_0)_{NF} = (\gamma_1/\gamma_{one})K^{-1}[r^3/R_{s,1}{}^3 - 2r^2\ell/R_{i,1}{}^3] + (\gamma_0/\gamma_{one})K^{-1}[r^3 R_{s,2}{}^3 - 2r^2\ell/R_{i,2}{}^3] \ . \tag{S70}$$

Now, for practical emitters $K \approx 1$. As above, we work in an approximation in which $(\gamma_1/\gamma_{one})$ and $(\gamma_0/\gamma_{one})$ are both set equal to 1. Hence we obtain:

$$(\delta_0)_{NF} = [r^3/R_{s,1}{}^3 - 2r^2\ell/R_{i,1}{}^3] + [r^3 R_{s,2}{}^3 - 2r^2\ell/R_{i,2}{}^3] \ . \tag{S71}$$

and, on substituting for the "R"s, using eqs (S1) and (S3):

$$(\delta_0)_{NF} \approx r^3/\{c^2+r^2\}^{3/2} - 2r^2\ell/\{c^2+(2\ell)^2\}^{3/2} + r^3/\{4c^2+r^2\}^{3/2} - 2r^2\ell/\{4c^2+(2\ell)^2\}^{3/2} \tag{S72}$$

When $r \ll c$, this reduces to:

$$(\delta_0)_{NF} \approx r^3/c^3 - 2r^2\ell/\{c^2+(2\ell)^2\}^{3/2} + (1/8)(r^3/c^3) - 2r^2\ell/\{4c^2+(2\ell)^2\}^{3/2}. \tag{S73}$$

For small separations ($r \ll c \ll \ell$), this reduces to

$$(\delta_0)_{NF} \approx r^3/c^3 - (1/4)r^2/\ell^2 + (1/8)(r^3/c^3) - (1/4)r^2/\ell^2 \ . \tag{S74}$$

$$(\delta_0)_{NF} \approx (9/8)(r/c)^3 - (1/2)(r/\ell)^2. \tag{S75}$$

**For sufficiently small separations [($r \ll c \ll \ell$) and ($c^3 \ll 2r\ell^2$) and ($c \geq 4r$)]**

$$(\delta_0)_{NF} \approx (9/8)(r/c)^3 = (9/8)\eta^3(\ell/c)^3 \tag{S76}$$

Since, from earlier, $(\delta_1)_{NF} \approx 2\eta^3(\ell/c)^3$, it follows from eq. (48) that

$$(\Xi_{NF}-1) \approx (\delta_0)_{NF} - (\delta_1)_{NF} = -(7/8)\eta^3(\ell/c)^3 \tag{S77}$$

**For large separations ($r \ll \ell \ll c$),** eq. (S73) reduces to:

$$(\delta_0)_{NF} \approx r^3/c^3 - 2r^2\ell/c^3 + (1/8)(r^3/c^3) - (1/4)\,r^2\ell/c^3 \tag{S78}$$

$$(\delta_0)_{NF} \approx (9/8)(r^3/c^3) - (9/4)r^2\ell/c^3 = (9/8)(r^2/c^3)(r-2\ell) \tag{S79}$$

Hence (because $r \ll \ell$)

$$(\delta_0)_{NF} \approx (r^2/c^3)[-(9/4)\ell] = -(9/4)\eta^2(\ell/c)^3 \tag{S80}$$

Since, from earlier, $(\delta_1)_{dir} \approx -4\eta^2(\ell/c)^3$, it follows from eq. (48) that

$$(\Xi_{NF}-1) \approx (\delta_0)_{NF} - (\delta_1)_{NF} = +(7/4)\eta^2(\ell/c)^3 \ . \tag{S81}$$

Theseå results provide the bottom line in Table V.

\*\*\*\*\*\*\*\*\*\*\*\*\*\*\*\*\*\*\*\*